# Enhancing Aviation Communication Transcription: Fine-Tuning Distil-Whisper with LoRA

Shokoufeh Mirzaei,[*]
*California State Polytechnic University, Pomona, California, 91768, USA*

Jesse Arzate.[†]
*Edwards Air Force Base Test Center, California, 93524, USA*

Yukti Vijay[‡]
*Skyryse, San Fransisco, California, 94103, USA*

**Transcription of aviation communications has several applications, from assisting air traffic controllers in identifying the accuracy of read-back errors to search and rescue operations. Recent advances in artificial intelligence have provided unprecedented opportunities for improving aviation communication transcription tasks. OpenAI's Whisper is one of the leading automatic speech recognition models. However, fine-tuning Whisper for aviation communication transcription is not computationally efficient. Thus, this paper aims to use a Parameter-Efficient Fine-tuning method called Low-Rank Adaptation to fine-tune a more computationally efficient version of Whisper, distil-Whisper. To perform the fine-tuning, we used the Air Traffic Control Corpus dataset from the Linguistic Data Consortium, which contains approximately 70 hours of controller and pilot transmissions near three major airports in the US. The objective was to reduce the word error rate to enhance accuracy in the transcription of aviation communication. First, starting with an initial set of hyperparameters for LoRA (Alpha = 64 and Rank = 32), we performed a grid search. We applied a 5-fold cross-validation to find the best combination of distil-Whisper hyperparameters. Then, we fine-tuned the model for LoRA hyperparameters, achieving an impressive average word error rate of 3.86% across five folds. This result highlights the model's potential for use in the cockpit.**

---

[*] Professor and Chair, Industrial and Manufacturing Engineering Department
[†] Data Scientist, Edwards Air Force Base
[‡] Flight Systems Test Engineer, Skyryse

## Nomenclature

| | | |
|---|---|---|
| *S* | = | the number of substitution errors in transcription |
| *D* | = | the number of deletion errors in transcription |
| *I* | = | the number of insertion errors in transcription |
| *N* | = | the total number of words in the reference transcription |

## Introduction

Radio communications are a critical air traffic control system as they build a strong bond between the pilot and controller and other manned/unmanned aerial vehicles. The single most important principle in pilot-controller communications is understanding. In aviation communication, while brevity is critical, and contacts must be kept as brief as possible, pilots and controllers must still effectively fulfill their control responsibilities. Since concise phraseology may not always suffice, pilots might deviate from standard phrasing and use alternative words to convey their message, particularly in high-stress situations. Having a reliable transcription system that can assist in understating and verifying the audio message received can improve the safety and quality of such communications. In aviation-related tasks, the errors of the automatic speech recognition (ASR) system can compromise flight safety. Therefore, developing a robust ASR with minimum error is essential.

The literature includes studies investigating the potential applications of ASR techniques in the Air Traffic Control (ATC) domain, including safety monitoring of live operations [1–4]. Researchers have used several open-source ASR systems. Examples are Bavieca [5], CMU Sphinx [6], HMM toolkit [7] Kaldi [8], Deep Speech [9], wav2letter++ [10], and Jasper [11]. Whisper, Developed by Open AI in late 2022 [12], stands at the forefront of automatic speech recognition models. It aimed to enhance the accuracy of transcribing audio into written text. The Whisper model is trained on either English-only or multilingual data using large-scale weak supervision. The English-only model is trained on speech recognition, in which the model predicts transcriptions in the same language as the audio. A key benefit of using Whisper models over classical approaches such as Hidden Markov Models-based models [5-8] is the ease of training as they do not require complicated pipelines with extensively engineered processing stages.

There are two ways approaches to using pre-trained models such as Whisper for the aviation communication transcription tasks: 1) training it from scratch or 2) fine-tuning the model for the specific task. Training a deep neural network from scratch requires extensive labeled data and computational power. For example, training Whisper from scratch for the specific task of transcribing pilot-ATC communications requires a large audio dataset along with their transcription. Creating such a dataset is particularly difficult due to the sensitive nature of ATC audio recordings and the domain knowledge needed to label the data. As a result, training Whisper from scratch might not be a feasible option in resources-constraint environment, which makes fine-tuning a more viable option. While it may be more feasible in comparison to training from scratch, the fine-tuning of ASR models involves updating all its parameters (also knowns as weights), which is time-consuming and computationally expensive. As the size of pre-trained models increases, computational resource requirements grow extensively. For example, T5 [13] has 770 million parameters, seven times more than BERT [14], which has 110 million parameters. Due to the inefficiency of pre-trained models in fine-tuning, parameter-efficient fine-tuning (PEFT) methods have emerged. These methods offer an effective solution by reducing the number of fine-tuning parameters and memory usage without significantly compromising the performance of the original fine-tuning techniques. Since 2019, several PEFT methods have been introduced. Xu et al. [15] categorized these methods into additive fine-tuning, reparametrized fine-tuning, hybrid fine-tuning, and partial fine-tuning. Among these categories, reparametrized fine-tunning methods such as Low-Rank Adaptation (LoRA) [16] have received noticeable attention in the literature for their computational efficiency. These methods reduce the number of trainable parameters by utilizing low-rank transformation while allowing operations with high-dimensional matrices (e.g., pre-trained weights).

In this paper, we use a two-pronged approach to tackle the computational challenges of fine-tunning Whisper: first, we use a more computationally effective version of Whisper called distil-Whisper [17]. Second, we used LoRA to fine-tune distil-Whisper. The closest work to this paper is the paper by Arra et al. [18], where Whisper is fine-tuned with LoRA for ATC audio transcription. However, in their work, the Whisper model was trained on publicly available data from a wide range of ATC sources with a thick European accent, which is not representative of US aviation communication. Also, their fine-tunning task only involved changing LoRA parameters (Alpha and Rank) and they did not perform cross-validation of their model. Thus, the contribution of this paper, which is also the distinction of this work from Arra et al [18], is three-fold: 1) Starting with an initial LoRA's hyperparameters, five-fold cross-validation was performed to find the best set of hyperparameters for fine-tunning distil-Whisper; 2) Using the best distil-Whisper hyperparameters, another five-fold cross-validation was performed to refine LoRA's hyperparameters for further fine-tuning of distil-Whisper. Additionally, the dataset used in this research, Air Traffic Control Corpus (ATC0) from the Linguistic Data Consortium, contains ATC communications from three US airports [19].

Section 2 will discuss the method used for fine-tuning distil-Whisper with LoRA. This section includes the dataset used, the process for cleaning the dataset, the performance metric, and the description of fine-tunning tasks. Section 3 will provide a conclusion and recommendation for future work.

## Methodology

### A. Data Set

The data for this study is collected from the Air Traffic Control Corpus (ATC0) from the Linguistic Data Consortium [19]. Texas Instruments collected the ATC0 Corpus under contract to DARPA. It was produced by the National Institute of Standards and Technology and distributed by the Linguistic Data Consortium. The ATC0 comprises recorded speech to support research and development activities in robust speech recognition domains similar to air traffic control (several speakers, noisy channels, relatively small vocabulary, constrained language, etc.). The audio data comprises voice communication traffic between various controllers and pilots. The audio files are 8kHz, 16-bit linear sampled data, representing continuous monitoring of a single Federal Aviation Administration (FAA) frequency for one to two hours without squelch or silence elimination. Some files indicate the amplitude of the received amplitude modulation (AM) carrier signal at 10 millisecond intervals. There are also transcription text files corresponding to each audio file. The transcriptions include the start and end times of each transmission, and each flight is identified by its flight number. The ATC0 consists of three sub-corpora from three airports in which the transmissions were collected -- Dallas Fort Worth (DFW), Logan International (BOS), and Washington National (DCA). The complete set contains approximately 70 hours of controller and pilot transmissions collected via antennas and radio receivers near the airports.

Whisper performs best with ingesting audio files of size 25MB or less. Thus, to curate the dataset for fine-tuning tasks, the 52 audio files included in the ATC0 dataset were segmented by start and end time of each line in the transcription text file so that each audio segment is between 5 and 30 seconds long. Each line in the transcript formed a data point that was then linked to a start time, end time and the designated audio segment. The segments were then converted from 8kHz to 16kHz to match Whisper's desired sampling rate. Case was standardized amongst the text files, and numeric characters (1,2,220) were converted to word representations such as one, two and two hundred and twenty. Each data point containing Rows of "UNINTELLIGIBLE" text was dropped from the dataset since training a model with "UNINTELLIGIBLE" could skew the word error rate (WER). The dataset contained whitespace before and after many words and the beginning and end of strings; all this whitespace was erased. By doing these steps, a dataset of 29091 data points was cleaned and ready for the fine-tuning tasks.

**B. Performance Metric**

We evaluated the performance of the ASR model using WER. It is one of the most common metrics used in ASR tasks. It is the ratio of the total number of errors in the transcription to the total number of words in the reference transcription. The errors can be classified into three types: 1) Substitutions: A substitution error occurs when a word in the transcribed output differs from the corresponding word in the reference transcription; 2) Deletions: A deletion error occurs when a word in the reference transcription is missing from the transcribed output; and 3) Insertions: An insertion error occurs when a word is present in the transcribed output that is not in the reference transcription. The WER score is calculated using Equation (1):

$$WER = (S + D + I) / N \qquad (1)$$

For example, if "The quick brown fox jumps over the lazy dog" is transcribed to "The quick brown dog jumps over the lazy," there is one substitution error where "fox" was incorrectly transcribed as "dog," one deletion error where the word "dog" was omitted, and no insertion errors. This results in a total of two errors out of nine words yielding a WER of 0.22 or 22%.

A lower WER score indicates higher accuracy, with a WER score of 0 indicating perfect accuracy. A WER score of less than 10% is considered suitable for many ASR applications, while a WER score of less than 5% is considered excellent. However, the WER threshold can vary depending on the complexity of the audio input, the quality of the audio, and the specific use case of the ASR system. When using WER as a metric, words must be aligned properly; otherwise, the WER becomes inflated. For the calculation of WER, we used the "JiWER" Python package to consider word alignment in calculating WER [20].

In fine-tunning Whisper, loss refers to the difference between the model's predicted output and target transcription. Typically, Whisper is trained using the cross-entropy loss, which measures how well the predicted token probabilities match the ground truth tokens. The model minimizes this loss during training by adjusting its weights, improving transcription accuracy. A lower loss indicates that the model's predictions are closer to the target transcriptions, leading to better performance, often reflected in a lower WER [12]. In this paper, we will report both loss and WER and show the correlation between these metrics later in this section.

### C. Fine-tuning Distil-Whisper Hyperparameters

The pre-trained Whisper model's capabilities can be enhanced through fine-tuning. However, given the inherent complexity and size of ASR models like Whisper, the fine-tuning process tends to increase computational demands. A meticulous selection of hyperparameters, substantial computing resources, and potentially large amounts of domain-specific audio are required to achieve optimal results. Since running Whisper large models in resource-constrained environments is challenging, we used the distil-Whisper model in this paper. This model is 5.8 times faster than Whisper, with 49% fewer parameters, while providing comparable WER on out-of-distribution test data. The distil-Whisper is designed to be paired with Whisper for speculative decoding [17].

To further expedite the fine-tuning process in resource-constrained environments, we used a Parameter-Efficient Fine-Tunning method called Low-Rank Adoption (LoRA) [16]. It involves a low-rank decomposition of the weight matrix to decrease the overall number of trainable parameters. So, instead of directly modifying all the neural network weights, only a small low-rank subset of the model weights is revised. Decreasing the trainable parameters is crucial because pre-trained models, such as distil-Whisper, have millions of trainable model weights. Therefore, modifying all the weights at once is computationally expensive and time-consuming. Moreover, since the pre-trained models have already been trained to perform a particular task, by modifying these weights to improve its performance for a specific task, there is a chance that the new model may lose its capability to perform well on the original task, also known as catastrophic forgetting [18]. LoRA freezes the original layers and injects a new trainable low-rank weight matrix into each layer of the Transformer architecture to reduce the number of trainable parameters, required computational power, and the possibility of catastrophic forgetting of the newly trained model.

To perform the fine-tuning of distil-Whisper with LoRA, given LoRA's hyperparameters of Alpha =32 and Rank=64, we chose batch sizes of 6 and 12, learning rates of 1e-5, 3e-5, 5e-4, and the number of epochs of 3 and 5. The combination of parameters was chosen considering the computational resources available for the task. We had access to 20 DL160 compute nodes to conduct the fine-tuning, allowing us to conduct 20 fine-tuning tasks simultaneously. For the implementation of the 5-fold cross-validation, the data was divided into five distinct folds, and every time, four folds (80% of the data or 23,273 rows) were used for training, and 1-fold (20% or 5,818 rows) was kept for testing. Combining hyperparameters and the number of folds resulted in 60 fine-tuning tasks.

Table 1 presents WER for fine-tunning distil-Whisper on the ATC0 dataset across five folds, evaluating the impact of batch size, learning rate, and the number of epochs across all folds. Increasing the number of epochs from three to five results in a consistent reduction in WER. This trend is observed for all learning rates and batch sizes, suggesting that additional training improves model performance. The effect is particularly pronounced at lower learning rates, where the model benefits from prolonged exposure to the data to achieve convergence. This result is also depicted in Fig 1.

**Table 1 WER resulted from fine-tuning distil-Whisper with LoRA on ATC0 dataset**

| Fold | Batch Size | 6 | | 12 | |
|---|---|---|---|---|---|
| | **Learning Rate/Epoch** | **3** | **5** | **3** | **5** |
| 1 | 1.00E-05 | 44.72 | 39.04 | 49.90 | 43.35 |
| | 3.00E-05 | 32.99 | 29.07 | 37.16 | 32.12 |
| | 5.00E-04 | 7.27 | 3.21 | 9.22 | 3.85 |
| 2 | 1.00E-05 | 43.84 | 38.64 | 49.22 | 42.95 |
| | 3.00E-05 | 33.11 | 28.81 | 36.71 | 32.02 |
| | 5.00E-04 | 13.40 | 11.27 | 15.01 | 12.67 |
| 3 | 1.00E-05 | 43.45 | 38.07 | 48.81 | 42.71 |
| | 3.00E-05 | 31.68 | 27.03 | 35.81 | 31.08 |
| | 5.00E-04 | 7.01 | 3.26 | 8.73 | 4.23 |
| 4 | 1.00E-05 | 43.11 | 38.07 | 48.85 | 42.99 |
| | 3.00E-05 | 32.03 | 27.17 | 35.84 | 30.97 |
| | 5.00E-04 | 7.19 | 3.19 | 8.98 | 4.22 |
| 5 | 1.00E-05 | 46.68 | 41.44 | 52.70 | 46.13 |
| | 3.00E-05 | 34.92 | 30.48 | 39.25 | 34.03 |
| | 5.00E-04 | 8.97 | 5.06 | 12.06 | 6.31 |

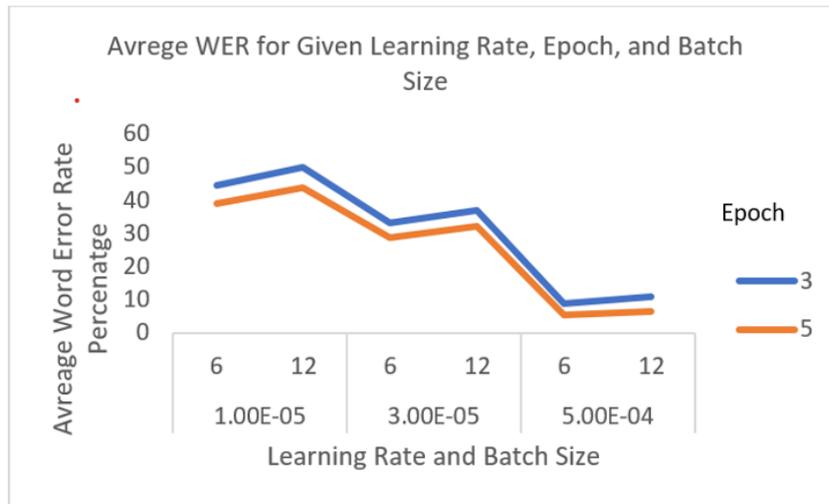

**Fig. 1 Average WER vs. learning rates, epochs, and batch Sizes**

The learning rate significantly influences WER. The lowest learning rate, 1e-5, results in the highest error rates across all folds and batch sizes, indicating slow learning and suboptimal convergence. The intermediate learning rate, 3e-5, substantially reduces WER, demonstrating a more balanced trade-off between learning speed and stability. The highest learning rate, 5e-4, yields the lowest WER values, particularly when trained for five epochs. However, in some cases, such as Fold 2, the error rate increases slightly, suggesting a degree of instability at this learning rate.

Batch size exerts a measurable effect on WER. Across all learning rates and epochs, the model trained with a batch size of 6 consistently achieved a lower WER than a batch size of 12. This outcome suggests that a smaller batch size allows for more stable parameter updates and improved generalization, potentially mitigating the risk of overfitting or poor convergence associated with larger batches.

The lowest WER values are obtained with a learning rate of 5e-4, five epochs, and a batch size of 6, with Fold 4 yielding the best result at 3.19. Conversely, the highest error rates occur at a learning rate 1e-5 with three epochs and batch size 12, reaching 52.70 in Fold 5. These results indicate that fine-tunning distil-Whisper with LoRA benefits from an aggressive learning rate and a smaller batch size when trained for sufficient epochs.

Table 2 presents the loss values for fine-tuning distil-Whisper with LoRA on the ATC0 dataset across five folds, evaluating the effects of batch size, learning rate, and the number of training epochs.

**Table 2 Loss resulted from fine-tuning distil-Whisper with LoRA on ATC0 dataset**

| Fold | Batch Size | 6 | | 12 | |
|---|---|---|---|---|---|
| | Learning Rate/Epoch | 3 | 5 | 3 | 5 |
| 1 | 1.00E-05 | 1.43 | 1.14 | 1.75 | 1.39 |
| | 3.00E-05 | 0.92 | 0.71 | 1.14 | 0.89 |
| | 5.00E-04 | 0.29 | 0.21 | 0.35 | 0.25 |
| 2 | 1.00E-05 | 1.43 | 1.14 | 1.74 | 1.40 |
| | 3.00E-05 | 0.92 | 0.71 | 1.14 | 0.89 |
| | 5.00E-04 | 0.30 | 0.21 | 0.35 | 0.25 |
| 3 | 1.00E-05 | 1.42 | 1.14 | 1.74 | 1.40 |
| | 3.00E-05 | 0.92 | 0.71 | 1.13 | 0.89 |
| | 5.00E-04 | 0.30 | 0.21 | 0.35 | 0.25 |
| 4 | 1.00E-05 | 1.41 | 1.14 | 1.75 | 1.40 |
| | 3.00E-05 | 0.91 | 0.71 | 1.14 | 0.88 |
| | 5.00E-04 | 0.30 | 0.21 | 0.35 | 0.25 |
| 5 | 1.00E-05 | 1.56 | 1.26 | 1.91 | 1.54 |
| | 3.00E-05 | 1.02 | 0.79 | 1.27 | 0.99 |
| | 5.00E-04 | 0.33 | 0.23 | 0.40 | 0.28 |

The data presented in Table 2 follows the same trends observed in Table 1, confirming the effects of batch size, learning rate, and number of epochs on model performance. Increasing the number of epochs reduces loss consistently, mirroring the WER improvements. Learning rate plays a crucial role, with 1e-5 leading to the highest loss, 3e-5 providing a balanced reduction, and 5e-4 achieving the lowest values, although with slight instability in some cases, particularly in Fold 5. This trend is also shown in Figure 2.

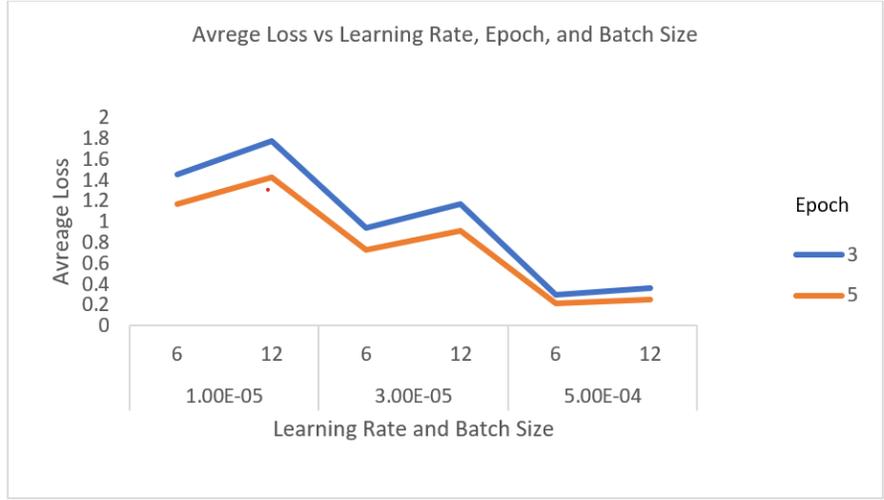

**Fig. 2 Average loss vs. learning rates, epochs, and batch Sizes**

Batch size influences loss similarly to WER, with smaller batches consistently yielding better results. The lowest loss, 0.21, occurs at 5e-4 with five epochs and batch size 6, while the highest loss, 1.91, is recorded at 1e-5 with three epochs and batch size 12. These findings reinforce the earlier conclusions, indicating that the optimal settings for reducing WER minimize loss.

Table 3 presents the average training time across five folds for fine-tuning distil-Whisper with LoRA, considering different batch sizes, learning rates, and epoch counts. As expected, training time increases with the number of epochs, with five epochs consistently requiring more time than three across all settings. Batch size has a notable impact on training duration. For the same learning rate and epoch configuration, a batch size of 6 leads to longer training times than a batch size of 12. This longer training time is because smaller batch sizes require more iterations per epoch, leading to increased computation time despite potential benefits in convergence.

**Table 3 Average training time across five folds for fine-tuning distil-Whisper with LoRA**

| Batch Size | 6 | | 12 | |
|---|---|---|---|---|
| Learning Rate/Epoch | 3 | 5 | 3 | 5 |
| 1.00E-05 | 203.48 | 279.54 | 179.24 | 263.73 |
| 3.00E-05 | 187.65 | 268.24 | 153.28 | 242.74 |
| 5.00E-04 | 159.06 | 271.66 | 164.78 | 268.33 |

The learning rate also influences training duration. Higher learning rates tend to reduce training time, particularly with three epochs, as seen in the decrease from 203.48 hours (1e-5) to 159.06 hours (5e-4) for batch size 6. However, this pattern is less consistent at five epochs, likely due to variations in convergence behavior and computational overhead from more significant updates. Overall, the shortest training time (153.28 minutes) occurs at batch size 12, learning rate 3e-5, and three epochs, while the longest (279.54 minutes) is recorded at batch size 6, learning rate 1e-5, and five epochs. As shown in Fig. 3, these results suggest that increasing batch size and learning rate can improve training efficiency, although with potential trade-offs in model performance.

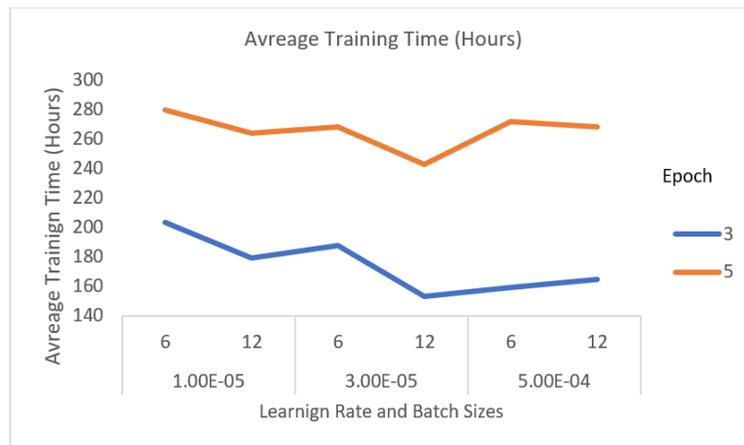

Fig. 3 Average training time (hours) vs. learning rates, epochs, and batch Sizes

D. **Fine-tuning LoRA Hyperparameters**

In the previous section, we concluded that fine-tuning distil-Whisper with LoRA benefits significantly from higher learning rates (5e-04), smaller batch sizes (6), and more extended training (5 epochs). These configurations resulted in the lowest WER and loss, making them the most effective for improving transcription accuracy. Building on this insight, we refine LoRA's hyperparameters to enhance model performance. LoRA introduces low-rank matrices to adapt a pre-trained model to a new task without significantly modifying the original model weights. In LoRA, Alpha (α) and Rank (r) are key hyperparameters controlling the extent of adaptation. Alpha (Scaling Factor) is the parameter that determines how much the LoRA updates contribute to the final weight adjustments. A higher Alpha increases the influence of the LoRA-adapted weights on the model's predictions, potentially leading to better learning and a higher risk of overfitting if set too high. Rank (r) controls the dimensionality of the low-rank matrices added to the model. A higher rank allows for greater flexibility in capturing new patterns but also introduces more parameters, increasing computational cost and the risk of instability. Conversely, a lower rank restricts adaptation but helps maintain generalization, making it a crucial trade-off in fine-tuning.

In this work, we evaluate the impact of two levels of Alpha (256 and 512) and two levels of Rank (256 and 512), leading to 20 fine-tuning experiments across five folds. The results, summarized in Table 4, show how these parameters influence WER.

**Table 4 Average WER for LoRA's Alpha and Rank Hyperparemetes across five folds**

| Alpha/ Rank | Rank= 256 | Rank= 512 |
|---|---|---|
| Alpha= 256 | 3.86 | 3.86 |
| Alpha= 512 | 22.44 | 38.45 |

The results indicate that when Alpha is set to 256, the WER average remains consistently low at 3.86% regardless of whether Rank is 256 or 512. However, when Alpha is increased to 512, the WER rises significantly, reaching a WER of 22.44% for Rank = 256 and further escalating to 38.45% for Rank = 512. This trend suggests that increasing Alpha beyond 256 negatively impacts transcription accuracy, leading to significantly higher WER. The stability in WER at Alpha = 256 implies that a lower Alpha value is more effective for maintaining performance while increasing Rank alone does not degrade WER. However, increasing Rank amplifies the negative impact on accuracy at higher Alpha values. These findings show the importance of carefully tuning Alpha, as excessive values can lead to degraded fine-tuning performance when adapting distil-Whisper with LoRA.

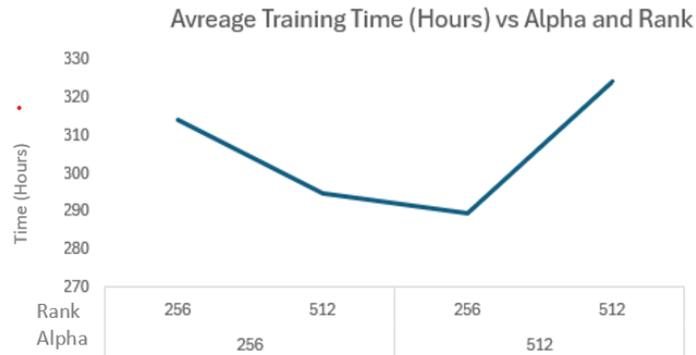

**Fig. 4 Average training hours vs. LoRA's hyperparameters, Alpha and Rank**

Fig. 4 shows Rank and Alpha's influence on distil-Whisper's average fine-tuning time. Contrary to one's expectations, increasing Rank does not consistently increase training time. At an alpha of 256, raising the Rank from 256 to 512 decreases training time. Conversely, at an alpha of 512, increasing Rank from 256 to 512 increases training time, aligning with the expected behavior due to the larger number of trainable parameters. This long training time suggests a strong interaction effect between Alpha and Rank.

The influence of Alpha also exhibits context-dependent behavior. At a rank of 256, increasing Alpha from 256 to 512 decreases training time, potentially indicating faster convergence. However, at a rank of 512, the same alpha increase increases training time, contradicting the previous trend. This flip in the alpha effect based on the rank value underscores the complex interplay between these two hyperparameters. The observed inconsistencies challenge the simplistic assumption that a higher Rank invariably leads to more extended training and that Alpha consistently scales convergence speed.

## Conclusion

This paper presents the fine-tuning of distil-Whisper using LoRA to transcribe aviation communication, utilizing approximately 70 hours of controller and pilot transmissions collected near three US airports. Initial fine-tuning experiments revealed a lower WER and loss were obtained with smaller batch sizes, higher learning rates, and increased epochs, even though these configurations required longer training times. The best configuration was achieved with a learning rate of 5e-04, 5 epochs, and a batch size of 6, resulting in the lowest WER and loss, demonstrating its effectiveness for aviation communication taks. Further fine-tuning of LoRA's hyperparameters—Rank and Alpha—enhanced transcription accuracy. The experiments showed complex interactions between hyperparameters of LoRA. An Alpha of 256 consistently yielded the lowest WER across different Rank values. However, increasing Alpha to 512 significantly reduced accuracy, emphasizing the importance of careful hyperparameter selection. The combination of Alpha = 256 with Rank = 256 or 512 achieved the lowest average WER of 3.86%, illustrating LoRA's capability to enhance fine-tuning efficiency without compromising accuracy.

Analysis of training time showed that increasing Rank generally led to higher computational costs and longer training times. However, the extent of this increase varied with the chosen Alpha. Notably, at Alpha = 256, the training time difference between Rank = 256 and Rank = 512 was smaller than at Alpha = 512, indicating an interaction effect. Although training duration varied, the substantial improvements in transcription accuracy, particularly at Alpha = 256, justified the computational cost, given the high accuracy requirements for cockpit communication.

Future work will include out-of-distribution testing to evaluate the model's generalization capability beyond the training data. This will assess the robustness of the fine-tuned model in real-world aviation communication scenarios, ensuring its reliability under varying operational conditions.

**Funding Sources**

This research was supported in part by the Air Force Research Laboratory Edwards Air Force Base Test Center Directorate, through the Air Force Office of Scientific Research Summer Faculty Fellowship Program®, Contract Numbers FA8750-15-3-6003, FA9550-15-0001 and FA9550-20-F-0005.

**References**

[1] Chen, S., Kopald, H. D., Chong, R., Wei, D., and Levonian, Z., "Read Back Error Detection using Automatic Speech Recognition," 2017. Available: https://www.semanticscholar.org/paper/Read-Back-Error-Detection-using-Automatic-Speech-Chen-Kopald/64018f82ee972144b46deb3073295f58010bfcff.

[2] Subramanian, S. V., Kostiuk, P. F., and Katz, G., "Custom IBM Watson Speech-to-Text Model for Anomaly Detection using ATC-Pilot Voice Communication," *2018 Aviation Technology, Integration, and Operations Conference*, American Institute of Aeronautics and Astronautics, 2018. https://doi.org/10.2514/6.2018-3979.

[3] Lin, Y., Deng, L., Chen, Z., Wu, X., Zhang, J., and Yang, B., "A Real-Time ATC Safety Monitoring Framework Using a Deep Learning Approach," *IEEE Transactions on Intelligent Transportation Systems*, Vol. 21, No. 11, 2020, pp. 4572–4581. https://doi.org/10.1109/TITS.2019.2940992.

[4] Badrinath, S., and Balakrishnan, H., "Automatic Speech Recognition for Air Traffic Control Communications," *Transportation Research Record*, Vol. 2676, No. 1, 2022, pp. 798–810. https://doi.org/10.1177/03611981211036359.

[5] Bolaños, D., "The Bavieca open-source speech recognition toolkit," *2012 IEEE Spoken Language Technology Workshop (SLT)*, Miami, FL, USA, 2012, pp. 354-359, https://doi.org/10.1109/SLT.2012.6424249.

[6] Lamere, P., Kwok, P., Gouvêa, E., Raj, B., Singh, R., Walker, W., et al., "The CMU SPHINX-4 Speech Recognition System," 2001. Available: https://www.semanticscholar.org/paper/THE-CMU-SPHINX-4-SPEECH-RECOGNITION-SYSTEM-Lamere-Kwok/5064c602c3a57f4e6f1e4c8f8fb137384c5d41a7.

[7] "HTK Speech Recognition Toolkit," 2020. Available: https://htk.eng.cam.ac.uk/.

[8] Povey, D., Ghoshal, A., Boulianne, G., Burget, L., Glembek, O., Goel, N., et al., "The Kaldi Speech Recognition Toolkit," *IEEE 2011 Workshop on Automatic Speech Recognition and Understanding*, IEEE Signal Processing Society, 2011.

[9] Amodei, D., Ananthanarayanan, S., Anubhai, R., Bai, J., Battenberg, E., Case, C., et al., "Deep Speech 2: End-to-End Speech Recognition in English and Mandarin," *Proceedings of the 33rd International Conference on Machine Learning*, PMLR, 2016, pp. 173–182. Available: https://proceedings.mlr.press/v48/amodei16.html.

[10] Pratap, V., Hannun, A., Xu, Q., Cai, J., Kahn, J., Synnaeve, G., et al., "Wav2Letter++: A Fast Open-Source Speech Recognition System," *ICASSP 2019 - IEEE International Conference on Acoustics, Speech and Signal Processing*, 2019, pp. 6460–6464. https://doi.org/10.1109/ICASSP.2019.8683535.

[11] Li, J., Lavrukhin, V., Ginsburg, B., Leary, R., Kuchaiev, O., Cohen, J. M., et al., "Jasper: An End-to-End Convolutional Neural Acoustic Model," *Interspeech 2019*, 2019, pp. 71–75. https://doi.org/10.21437/Interspeech.2019-1819.

[12] Radford, A., Kim, J. W., Xu, T., Brockman, G., Mcleavey, C., and Sutskever, I., "Robust Speech Recognition via Large-Scale Weak Supervision," *Proceedings of the 40th International Conference on Machine Learning*, PMLR, 2023, pp. 28492–28518. Available: https://proceedings.mlr.press/v202/radford23a.html.

[13] Raffel, C., Shazeer, N., Roberts, A., Lee, K., Narang, S., Matena, M., et al., "Exploring the Limits of Transfer Learning with a Unified Text-to-Text Transformer," arXiv:1910.10683, 2020. https://doi.org/10.48550/arXiv.1910.10683.

[14] Devlin, J., Chang, M.-W., Lee, K., and Toutanova, K., "BERT: Pre-training of Deep Bidirectional Transformers for Language Understanding," arXiv:1810.04805, 2019. https://doi.org/10.48550/arXiv.1810.04805.

[15] Ding, N., Qin, Y., Yang, G., Wei, F., Yang, Z., Su, Y., et al., "Parameter-Efficient Fine-Tuning of Large-Scale Pre-Trained Language Models," *Nature Machine Intelligence*, Vol. 5, No. 3, 2023, pp. 220–235. https://doi.org/10.1038/s42256-023-00626-4.

[16] Hu, E. J., Shen, Y., Wallis, P., Allen-Zhu, Z., Li, Y., Wang, S., et al., "LoRA: Low-Rank Adaptation of Large Language Models," arXiv:2106.09685, 2021. https://doi.org/10.48550/arXiv.2106.09685.

[17] Gandhi, S., von Platen, P., and Rush, A. M., "Distil-Whisper: Robust Knowledge Distillation via Large-Scale Pseudo Labelling," arXiv:2311.00430, 2023. https://doi.org/10.48550/arXiv.2311.00430.

[18] Arra, A., Achour, G., Payan, A., Harrison, E., and Mavris, D., "Automatic Speech Recognition Model Fine-Tuning and Development of a New Evaluation Metric for Terminal Airspace Safety Analysis," AIAA AVIATION FORUM AND ASCEND 2024, American Institute of Aeronautics and Astronautics, 2024. https://doi.org/10.2514/6.2024-4570.

[19] Linguistic Data Consortium, "Air Traffic Control Complete.", LDC94S14A. Available: https://catalog.ldc.upenn.edu/LDC94S14A.

[20] Leuven, W., "Measuring Word Error Rate in Python," *GitHub Repository*, 2020. Available: https://github.com/jitsi/jiwer.